\documentclass[prd,showpacs,preprintnumbers,amsmath,amssymb,nofootinbib]{revtex4}

\usepackage{graphics,color}
\usepackage{epsfig}
\usepackage{dcolumn}
\usepackage{bm}

\begin{document}
\title{Semileptonic decays of $B_{s1}$, $B_{s2}^*$, $B_{s0}$ and $B_{s1}'$}

\author{Zhi-Gang Luo$^1$}\author{Xiao-Lin Chen$^1$}\author{Xiang
Liu$^{1,2}$}\email{liuxiang@teor.fis.uc.pt}
\author{Shi-Lin Zhu$^1$}
\email{zhusl@phy.pku.edu.cn}

\affiliation{$^1$Department of Physics, Peking University, Beijing
100871, China \\ $^2$Centro de F\'{i}sica Te\'{o}rica,
Departamento de F\'{i}sica, Universidade de Coimbra, P-3004-516
Coimbra, Portugal}

\date{\today}

\begin{abstract}

Stimulated by recent observations of the excited bottom-strange
mesons $B_{s1}$ and $B_{s2}^*$, we calculate the semileptonic
decays $B_{s0},\,B_{s1}^{\prime},\,B_{s1},\,B_{s2}^*\to
[D_s(1968),\,D_{s}^*(2112),\,D_{sJ}(2317),\,
D_{sJ}(2460)]\ell\bar{\nu}$, which is relevant for the exploration
of the potential of searching these semileptonic decays in
experiment.

\end{abstract}


\maketitle

\section{introduction}\label{sec1}

Recently, the CDF collaboration announced the observation of two
orbitally excited narrow $B_s$ mesons. Their masses are
$m_{B_{s1}}=5829.4\pm0.7$ MeV and $m_{B_{s2}^*}=5839.6\pm 0.7$ MeV
\cite{CDF}. Later the D0 collaboration confirmed the $B_{s2}^{*}$
state with $m_{B_{s2}^*}=5839.6\pm 1.1(\mathrm{stat}.)\pm
0.7(\mathrm{syst}.)$ MeV \cite{D0}. Meanwhile, the D0
collaboration indicated that $B_{s1}$ was not observed with the
available data set \cite{D0}. In the heavy quark effective theory
(HQET), the charmed or bottom mesons can be categorized into
doublets since the angular momentum of the light components
$j_{\ell}$ is a good quantum number in the $m_Q\to\infty$ limit.
They are $j_{\ell}^P=\frac{1}{2}^-$, the $H$ doublet $(0^-, 1^-)$
with the orbital angular monument $L=0$,
$j_{\ell}^P=\frac{1}{2}^+$, the $S$ doublet $(0^+,1^+)$ and
$j_{\ell}^P=\frac{3}{2}^+$, the $T$ doublet $(1^+,2^+)$ with
$L=1$. The recently observed $B_{s1}$ and $B_{s2}^*$ mesons belong
to the $T$ doublet \cite{CDF,D0}.

The observations of $B_{s1}$ and $B_{s2}^*$ \cite{CDF,D0} not only
enrich the mass spectrum of the bottom-strange
system\footnote{Besides $B_{s1}$ and $B_{s2}^*$ observed by CDF
and D0, there are only two established bottom-strange states
($B_{s}$ and $B_s^*$) listed in Particle Data group (PDG) so far
\cite{PDG}.} but also inspire our interest in these two unobserved
P-wave bottom-strange mesons. Since $B_{s1}$ and $B_{s2}^*$ lie
above the thresholds of $BK$ and $B^*K$, their strong decay is
dominant. In our recent work \cite{liu-3P0}, we calculated the
strong decays of $B_{s1}$ and $B_{s2}^*$ with the $^3P_0$ model.
Our result indicates that the two-body strong decay widths of
$B_{s1}$ and $B_{s2}^*$ can reach up to 98 keV and 5 MeV,
respectively. In contrast, $B_{s0}$ and $B_{s1}'$ were generally
speculated to lie below the threshold of $BK$ and $B^*K$ in Ref.
\cite{mass-Bs01}. In fact, they are expected to be narrow
resonances with a width around several tens of keV \cite{lujie}
since their main decay modes are the isospin-violating strong
decays and electromagnetic decays. These two states are still
missing in the experiments. Up to now, they have not been observed
experimentally.

The semileptonic decays of
$B_{s0},\,B_{s1}^{\prime},\,B_{s1},\,B_{s2}^*$ have not been
explored. In this work, we will calculate the semileptonic decays
$B_{s0},\,B_{s1}^{\prime},\,B_{s1},\,B_{s2}^*\to
[D_s(1968),\,D_{s}^*(2112),\,D_{sJ}(2317),\,
D_{sJ}(2460)]\ell\bar{\nu}$ in the framework of the
constituent-quark meson (CQM) model, where
$(D_{s}(1968),D_{s}^*(2112))$ and $(D_{sJ}(2317),D_{sJ}(2460))$
belong to $H$ and $S$ doublets of the bottom-strange meson
respectively.

Through the present investigation, we want to learn the potential
of searching the semileptonic decays of $B_{s1}$, $B_{s2}^*$,
$B_{s0}$ and $B_{s1}'$ in future experiments such as CDF, D0 and
the forthcoming Large Hadron Collider beauty (LHCb). If these
semileptonic decays can be reached by the future experiments, one
may compare the bottom-strange mesons with these two exotic
charm-strange states $D_{sJ}(2317)$ and $D_{sJ}(2460)$ through the
semileptonic decays. The observations of the above narrow
charm-strange mesons have resulted in an extensive study of their
properties in the past five years\footnote{$D_{sJ}(2317)$ and
$D_{sJ}(2460)$ with spin parity structures $J^{P}=0^{+}$ and
$J^{P}=1^{+}$ \cite{2317,Belle,CLEO} have inspired heated debates
about their structures. A detailed review can be found in Ref.
\cite{review}. Possible interpretations include the $(0^+,1^+)$
chiral partners of $D_s$ and $D_s^*$ \cite{Bardeen}, P-wave
excited states of $D_s$ and $D_s^*$ \cite{Chao}, coupled-channel
effects between $c\bar s$ states and the $DK$ continuum
\cite{Beverenn1}, conventional $c\bar s$ states
\cite{dai0,Narison}, four-quark states \cite{Chen,Cheng,T. Barnes}
etc. Considering the large contribution of the S-wave $DK$
continuum in the QCD sum rule (QSR) approach, the mass of
$D_{sJ}^{*}(2317)$ agrees well with the experimental value
\cite{Dai}. Therefore, $D_{sJ}^*(2317)$ and $D_{sJ}(2460)$ are
very probably conventional $c\bar{s}$ states with $J^P=0^+$ and
$J^P=1^+$ \cite{review}.}.

This work is organized as follows. After the introduction, we
briefly introduce the theoretical framework, i.e. the CQM model.
In Section \ref{sec3}, the formulation relevant to the
semileptonic decays of $B_{s1}$, $B_{s2}^*$, $B_{s0}$ and
$B_{s1}'$ will be given. In Section \ref{sec4}, we present the
numerical results with all input parameters. The last section is a
short discussion. The detailed expressions are collected in the
appendix.

\section{The CQM model}\label{sec2}

We first give a brief review of the CQM model \cite
{CQM,feldmann}. The effective Lagrangian of the CQM model
incorporates both the heavy-quark spin-flavor symmetry and the
chiral symmetry \cite{CQM}
\begin{eqnarray}
\mathcal{L}_{CQM}&=&\bar{\chi}[\gamma\cdot(i\partial+\mathcal{V})]\chi+\bar{\chi}\gamma\cdot
\mathcal{A}\gamma_5\chi-m_q\bar{\chi}\chi
+\frac{f^2_{\pi}}{8}Tr[\partial^{\mu}\Sigma\partial_{\mu}\Sigma^+]+
\bar{h}_{v}(iv\cdot\partial)h_v\nonumber\\
&&-[\bar{\chi}(\bar{H}+\bar{S}+i\bar{T}^{\mu}\frac{\partial_{\mu}}{\Lambda})h_{v}+h.c.]
 +\frac{1}{2G_3}
Tr[(\bar{H}+\bar{S})(H-S)]+\frac{1}{2G_4}Tr[\bar{T}^{\mu}T_{\mu}].
 \label{cqm}
\end{eqnarray}
Here $H$, $S$ and $T$ denote the super-fields corresponding to the
($0^-,1^-$), ($0^+,1^+$) and ($1^+,2^+$) doublets respectively,
whose explicit matrix representations are \cite{luke}
\begin{eqnarray}
H&=&\frac{1+ v\!\!\!\slash}{2}[P_{\mu}^*\gamma^{\mu}-P\gamma_5],\\
S&=&\frac{1+v\!\!\!\slash}{2}[{P_{1}'}_{\mu}^{*}\gamma^{\mu}\gamma_5-P_{0}],\\
T^{\mu}&=&\frac{1+
v\!\!\!\slash}{2}\bigg\{P_2^{*\mu\nu}\gamma_{\nu}-\sqrt{\frac{3}{2}}P_{1\nu}^*\gamma_5\big[g^{\mu\nu}-
\frac{1}{3}\gamma^{\nu}(\gamma^{\mu}-v^{\mu})\big]\bigg\}.
\end{eqnarray}
$P$, $P^{*\mu}$, $P_{0}$ and ${P_{1}'}^{*}$ correspond to the
annihilation operators of the pseudoscalar, vector, scalar and
axial vector mesons respectively. They are normalized as
\begin{eqnarray*}
\langle0|P|M(0^-)\rangle&=&\sqrt{M_{H}},\quad\quad\quad
\langle0|P^{*\mu}|M(1^-)\rangle=\sqrt{M_{H}}\epsilon^{\mu},\\
\langle0|P_{0}|M(0^+)\rangle&=&\sqrt{M_{S}},\quad\quad\quad
\langle0|{P_{1}'}^{*\mu}|M(1^+)\rangle=\sqrt{M_{S}}\epsilon^{\mu},\\
\langle0|P_1^{*\mu}|M(1^+)\rangle&=&\sqrt{M_T}\epsilon^\mu,\quad\quad
\langle0|P_2^{*\mu\nu}|M(2^+)\rangle=\sqrt{M_T}\eta^{\mu\nu}.
\end{eqnarray*}
In Eq. (\ref{cqm}), the fifth term represents the kinetic term of
heavy quarks with $\slash\!\!\!vh_{v}=h_v$. $\chi=\xi q (q=u,d,s)$
denotes the light-quark field with
$\xi=e^{\frac{i\mathcal{M}}{f_{\pi}}}$ and $\mathcal{M}$ is the
octet pseudoscalar matrix. $\mathcal{V}^{\mu}$ and
$\mathcal{A}^{\mu}$ are defined as
\begin{eqnarray}
&&\mathcal{V}^{\mu}=\frac{1}{2}(\xi^{\dag}\partial^{\mu}\xi+\xi\partial^{\mu}\xi^{\dag}),\\
&&\mathcal{A}^{\mu}=\frac{-i}{2}(\xi^{\dag}\partial^{\mu}\xi-\xi\partial^{\mu}\xi^{\dag}).
\end{eqnarray}

An important feature of the effective Lagrangian in Eq.
(\ref{cqm}) is that $\mathcal{L}_{CQM}$ describes the interaction
vertex of the heavy-meson with heavy and light quarks, which makes
the study of the phenomenology of heavy-meson physics at the quark
level feasible. The CQM model has been applied to study the heavy
meson phenomenology
\cite{application,xiangliu-0,xiangliu-1,xiangliu-2,parameter}. In
Ref. \cite{xiangliu-2}, the semileptonic decays $B_{s}\to
D_{sJ}(2317,2460)\ell\bar\nu$ have been studied in the CQM model.
In this work, we extend the same formalism to the semileptonic
decays of $B_{s1}$ and $B_{s2}^*$. The interested readers may also
consult the review paper of the CQM model in \cite{CQM}.

\section{The semileptonic decays of $B_{s1}$, $B_{s2}^*$, $B_{s0}$ and $B_{s1}'$}\label{sec3}

\begin{figure}[htb]
\begin{center}
\scalebox{0.7}{\includegraphics{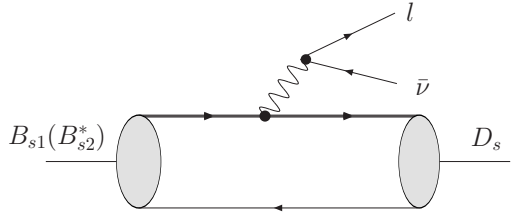}}
\end{center}
\caption{The Feynman diagram depicting the semileptonic decays of
$B_{s1}(B_{s2}^*)\rightarrow D_{s}\ell\bar{\nu}$. Here the
thick-line denotes the heavy-quark propagator.} \label{diagram}
\end{figure}

In this section, we calculate the semileptonic decays of $B_{s1}$,
$B_{s2}^*$, $B_{s0}$ and $B_{s1}'$ in the CQM model. We are
interested in the semileptonic decay modes of $B_{s1}$,
$B_{s2}^*$, $B_{s0}$ and $B_{s1}'$ including
$D_{s}(1968)\ell\bar\nu$, $D_{s}^*(2112)\ell\bar\nu$,
$D_{sJ}(2317)\ell\bar\nu$ and $D_{s}(2460)\ell\bar\nu$, which are
depicted in Fig. \ref{diagram}.

The four-fermion operator describing $b\to c+\ell\bar\nu$ is
\begin{eqnarray}
\mathcal{O}=\frac{G_FV_{cb}}{\sqrt{2}}\bar{c}\gamma^{\mu}(1-\gamma_5)b\bar{\nu}\gamma_{\mu}(1-\gamma_5)\ell.
\end{eqnarray}
Thus the general decay amplitudes of
$B_{s1}(B_{s2}^*,B_{s0},B_{s1}')\to
D_{s}(D_{s}^*,D_{sJ}(2317,2460))\ell\bar\nu$ read
\begin{eqnarray}
&&\mathcal{M}[(b\bar{s})\to
(c\bar{s})+\ell\bar\nu]\nonumber\\&&=\frac{G_FV_{cb}}{\sqrt{2}}\langle
(c\bar{s})|\bar{c}\gamma^{\mu}(1-\gamma_5)b|(b\bar{s})\rangle
\langle
\ell\bar{\nu}|\bar{\nu}\gamma_{\mu}(1-\gamma_5)\ell|0\rangle,\label{la}
\end{eqnarray}
where we use $(c\bar{s})$ and $(b\bar{s})$ to denote the
charm-strange and bottom-strange mesons, respectively. The
hadronic matrix element $\langle
(c\bar{s})|\bar{c}\gamma^{\mu}(1-\gamma_5)b|(b\bar{s})\rangle$ has
to be calculated using phenomenological models, such as QSR and
the CQM model. Here we adopt the CQM model to calculate the
hadronic matrix element. Figure \ref{diagram} illustrates the
semileptonic decays $B_{s1}(B_{s2}^*)\to D_{s}\ell\bar\nu$ in the
CQM model.

\begin{center}
\begin{table}[htb]
\begin{tabular}{c||c}\hline\hline
Interaction & Vertex \\\hline

$b-B_{s1}-\bar{s}$&
$(\frac{i}{\Lambda})\sqrt{\frac{3}{2}}\sqrt{m_{B_{s1}}Z_{T}}\;\frac{1+v\!\!\!\slash
}{2}\gamma_{5}[\epsilon\cdot k-\frac{\epsilon\!\!\!\slash}{3}
(k\!\!\!\slash-v\cdot k)]$\\

$b-B_{s2}^*-\bar{s}$&$(-\frac{i}{\Lambda})\sqrt{m_{B_{s2}^*}Z_{T}}\;\frac{1+v\!\!\!\slash
}{2}k_{\mu}\eta^{\mu\nu}\gamma_{\nu}$\\

$b-B_{s0}-\bar{s}$&$i\sqrt{m_{B_{s0}^*}Z_{S}}\;
\frac{1+v\!\!\!\slash }{2}$\\

$b-B_{s1}'-\bar{s}$&$-i\sqrt{m_{B_{s1}'}Z_{S}}\;
\epsilon\!\!\!\slash^*_2\gamma_{5}\;\frac{1+v\!\!\!\slash
}{2}$\\

$c-D_{s}-\bar{s}$&$-i\sqrt{m_{D_{s}}Z_{H}}\;\gamma_5
\frac{1+v\!\!\!\slash' }{2}$\\

$c-D_{s}^*-\bar{s}$&$-i\sqrt{m_{D_{s}}Z_{H}}\;\epsilon\!\!\!\slash^*_1
\frac{1+v\!\!\!\slash' }{2}$\\

$c-D_{s0}-\bar{s}$&$i\sqrt{m_{D_{s0}^*}Z_{S}}\;
\frac{1+v\!\!\!\slash' }{2}$\\

$c-D_{s1}'-\bar{s}$&$-i\sqrt{m_{D_{s1}'}Z_{S}}\;
\epsilon\!\!\!\slash^*_2\gamma_{5}\;\frac{1+v\!\!\!\slash'
}{2}$\\\hline\hline

\end{tabular}
\caption{The interaction vertex for the semileptonic decays of
$B_{s1}$, $B_{s2}^*$, $B_{s0}$ and $B_{s1}'$. \label{vertex}}
\end{table}
\end{center}

According to Eq. (\ref{cqm}), we give the interaction vertices in
Table \ref{vertex}, which are related to the semileptonic decays
of $B_{s1}$, $B_{s2}^*$, $B_{s0}$ and $B_{s1}'$. Here $\epsilon$,
$\eta$, $\epsilon_{1}$ and $\epsilon_{2}$ denote the polarization
vectors of the heavy mesons. Note that we use $D_{s0}$ and
$D_{s1}'$ to denote $D_{sJ}(2317)$ and $D_{sJ}(2460)$
respectively.  The normalization constants $Z_{H,S,T}$ are given
in Ref. \cite{CQM}:
\begin{eqnarray}
Z_{H}^{-1}&=&(\Delta_{H}+m_{s})\frac{\partial
\mathcal{I}_{3}(\Delta_{H})}{\partial\Delta_{H}}+\mathcal{I}_{3}(\Delta_{H}),\label{ZH}\\
Z_{S}^{-1}&=&(\Delta_{S}-m_{s})\frac{\partial
\mathcal{I}_{3}(\Delta_{S})}{\partial\Delta_{S}}+\mathcal{I}_{3}(\Delta_{S}),\label{ZS}\\
Z_{T}^{-1}&=&\frac{1}{3\Lambda^2}\bigg\{
(\Delta_{T}^2-m_{s}^2)[\mathcal{I}_{3}(\Delta_{T})+(m_{s}+\Delta_{T})\frac{\partial
\mathcal{I}_{3}(\Delta_{T})}{\partial\Delta_{T}}]\nonumber\\
&&+(m_{s}+\Delta_{T})[\mathcal{I}_{1}+2\Delta_{T}\mathcal{I}_{3}(\Delta_{T})+\frac{\partial
\mathcal{I}_{0}(\Delta_{T})}{\partial\Delta_{T}}]
+\mathcal{I}_{0}(\Delta_{T})+\Delta_{T}\mathcal{I}_{1}
\bigg\}\label{ZT}
\end{eqnarray}
with the definitions of $I_{0,1,3}$ listed in the appendix.

In terms of the above expressions, we can write the general
expression of the hadronic matrix element $\langle
(c\bar{s})|\bar{c}\gamma^{\mu}(1-\gamma_5)b|(b\bar{s})\rangle$ in
the CQM model
\begin{eqnarray}
&&\langle
(c\bar{s})|\bar{c}\gamma^{\mu}(1-\gamma_5)b|(b\bar{s})\rangle
\nonumber\\&&=-\frac{iN_{c}}{16\pi^4}\int \mathrm{d}^4 l \frac{
Tr[\gamma^{\mu}(1-\gamma_{5})\Gamma_{a}(\gamma\cdot
l+m_s)\Gamma_{b}]} {(l^2-m_{s}^2)(v\cdot l+\alpha)(v'\cdot
l+\beta)},\nonumber\\
\end{eqnarray}
where $\Gamma_{a}$ corresponds to the vertex of the interaction of
$B_{s1}(B_{s2}^*,B_{s0},B_{s1}')$ with $b$ and $\bar{s}$.
$\Gamma_{b}$ is the vertex describing the interaction of
charm-strange meson with $c$ and $\bar{s}$ quarks. $N_c$ denotes
the colors degrees of freedom and $N_{c}=3$.
$\alpha(\beta)=\Delta_{H,S,T}$ denotes the mass difference between
the heavy mesons and the heavy quark \cite{CQM}.

In the following, by substituting $\Gamma_{a}$ and  $\Gamma_{b}$
with the expression listed in Table \ref{vertex}, one obtains the
hadron matrix elements relevant to the semileptonic decays of
$B_{s1}$ and $B_{s2}^*$ with the transitions of
$B_{s1}(B_{s2}^*)\to D_{s0}(D_{s1}')$:

\begin{eqnarray}
&&\langle
D_{s0}(v')|\gamma^{\mu}(1-\gamma_5)|B_{s1}(v,\epsilon)\rangle\nonumber\\
\quad
&&=\sqrt{m_{B_{s1}}m_{D_{s0}}}\zeta(\omega)\Big[f\epsilon^{\mu}+(\epsilon\cdot
v')(h_{1}v^{\mu}+h_{2}{v'}^{\mu})
+ih_{3}\varepsilon^{\mu\delta\lambda\rho}v'_{\delta}v_{\lambda}\epsilon_{\rho}\Big],\label{1}
\end{eqnarray}
\begin{eqnarray}
&&\langle
D_{s1}'(v',\epsilon_2)|\gamma^{\mu}(1-\gamma_5)|B_{s1}(v,\epsilon)\rangle\nonumber\\
\quad&&=\sqrt{m_{B_{s1}}m_{D_{s1}'}}\zeta(\omega)\Big[f_{1}(v\cdot
\epsilon_2^*)
\epsilon^{\mu}+f_{2}(v'\cdot\epsilon)\epsilon_2^{*\mu}
+h_1(\epsilon_2^{*}\cdot
\epsilon)v^{\mu}+h_{2}(\epsilon_2^{*}\cdot
\epsilon){v'}^{\mu}+h_{2}'(v'\cdot \epsilon)(v\cdot
\epsilon_2^*){v'}^{\mu}\nonumber\\&&\quad+ih_3\varepsilon^{\mu\delta\lambda\rho}\epsilon_{2\delta}^*v_{\lambda}\epsilon_{\rho}
+ih_4\varepsilon^{\mu\delta\lambda\rho}\epsilon_{2\delta}^*v_{\lambda}'\epsilon_{\rho}+ih(v'\cdot\epsilon)
\varepsilon^{\mu\delta\lambda\rho}
\epsilon_{2\delta}^{*}v_{\lambda}'v_{\rho}\Big]
\end{eqnarray}
for the semileptonic decays of $B_{s1}$,

\begin{eqnarray}
&&\langle
D_{s0}(v')|\gamma^{\mu}(1-\gamma_5)|B_{s2}^*(v,\eta)\rangle\nonumber\\
\quad&&=\sqrt{m_{B_{s2}^*}m_{D_{s0}}}\zeta(\omega)\Big[f{v'}^{\alpha}g^{\beta\mu}+g{v'}^{\alpha}v^{\mu}{v'}^{\beta}
+ih\varepsilon^{\beta\mu\delta\lambda}v_{\delta}v_{\lambda}'{v'}^{\alpha}\Big]\eta_{\alpha\beta},
\end{eqnarray}
\begin{eqnarray}
&&\langle
D_{s1}'(v',\epsilon_2)|\gamma^{\mu}(1-\gamma_5)|B_{s2}^*(v,\eta)\rangle\nonumber\\
\quad&&=\sqrt{m_{B_{s2}^*}m_{D_{s1}'}}\zeta(\omega)\Big[f{v'}^{\alpha}{v'}^{\beta}\epsilon_2^{*\mu}
+f_1
{v'}^{\alpha}\epsilon_2^{*\beta}v^{\mu}+f_2{v'}^{\alpha}\epsilon_2^{*\beta}{v'}^{\mu}\nonumber\\&&\quad+g\;g^{\beta\mu}
(\epsilon_2^*\cdot
v){v'}^{\alpha}+ih_1\varepsilon^{\beta\mu\delta\lambda}\epsilon_{2\delta}^{*}v_{\lambda}{v'}^{\alpha}
+ih_2\varepsilon^{\beta
\mu\delta\lambda}\epsilon_{2\delta}^{*}{v'}_{\lambda}{v'}^{\alpha}\Big]\eta_{\alpha\beta}\label{8}
\end{eqnarray}
for the semileptonic decays of $B_{s2}^*$. Here $\omega=v\cdot
v'$.

\begin{center}
\begin{table}[htb]
\begin{ruledtabular}
\begin{tabular}{c|ccccccccccc}
$B_{s1}\to$&$\sqrt{6}f$&$\sqrt{6}f_1$&$\sqrt{6}f_2$&$\sqrt{6}h_1$&$\sqrt{6}h_2$&$\sqrt{6}h_2'$&$\sqrt{6}h_3$&
$\sqrt{6}h_4$&$\sqrt{6}h$\\\hline

$D_{s}$&$-(\omega^2-1)$&-&-&$\omega-2$&$-3$&-&$-(\omega+1)$&-&-\\
$D_{s0}$&$-(\omega^2-1)$&-&-&$\omega+2$&$-3$&-&$-(\omega-1)$&-&-\\

$D_{s}^*$&-&$\omega+1$&$2(\omega+1)$&$-(\omega+1)$&$\omega+1$&$-3$&$\omega+1$&$-(\omega+1)$&$3$\\
$D_{s1}'$&-&$-(\omega-1)$&$2(\omega-1)$&$\omega-1$&$\omega-1$&$-3$&$-(\omega-1)$&$-(\omega-1)$&$3$\\
\end{tabular}
\caption{The coefficients relevant to the transitions of
$B_{s1}\to D_s,D_{s}^*,D_{s0},D_{s1}'$.\label{coefficient}}
\end{ruledtabular}
\end{table}
\end{center}

\begin{center}
\begin{table}[htb]
\begin{ruledtabular}
\begin{tabular}{c|cccccccccc}
$B_{s2}^*\to$&$f$&$f_1$&$f_2$&$g$&$h$&$h_1$&$h_2$\\\hline

$D_{s}$&$\omega+1$&-&-&$-1$&$1$&-&-\\
$D_{s0}$&$\omega-1$&-&-&$-1$&$1$&-&-\\

$D_{s}^*$&$-1$&$1$&$1$&$-1$&-&$1$&$1$\\
$D_{s1}'$&$-1$&$-1$&$1$&$1$&-&$-1$&$1$\\
\end{tabular}
\caption{The coefficients relevant to the transitions of
$B_{s2}^*\to D_s,D_{s}^*,D_{s0},D_{s1}'$. \label{coefficient1}}
\end{ruledtabular}
\end{table}
\end{center}

It is worth noting that the Lorentz structures of the matrix
elements $B_{s1}(B_{s2}^*)\to D_{s}$ and $B_{s1}(B_{s2}^*)\to
D_{s}^*$ are similar to those of the transitions of
$B_{s1}(B_{s2}^*)\to D_{s0}$ and $B_{s1}(B_{s2}^*)\to D_{s1}'$
respectively. For different semileptonic decay, the coefficients
$f_{i}$, $h_{i}^{(')}$, $g$, $h$ are different; they are given in
Tables \ref{coefficient}-\ref{coefficient1}. Meanwhile, the form
factor $\zeta(\omega)$ for $B_{s1}(B_{s2}^*)\to D_{s0}(D_{s1}')$
should be replaced by $\xi(\omega)$ for $B_{s1}(B_{s2}^*)\to
D_{s}(D_{s}^*)$. In the above expressions of the hadron element
matrices, there exist two independent form factors $\xi(\omega)$
and $\zeta(\omega)$:
\begin{eqnarray}
\xi(\omega)&=&\Lambda^{-1}\sqrt{Z_{T}Z_{H}}\big[\mathcal{A}_{2}+\mathcal{A}_3-\mathcal{B}_2m_s\big]_{\alpha=\Delta_{T},
\beta=\Delta_{H}},\nonumber\\
\end{eqnarray}
\begin{eqnarray}
\zeta(\omega)&=&\Lambda^{-1}\sqrt{Z_{T}Z_{S}}\big[\mathcal{A}_{2}-\mathcal{A}_3+\mathcal{B}_2m_s\big]_{\alpha=\Delta_{T},
\beta=\Delta_{S}}.\nonumber\\
\end{eqnarray}

Equations (\ref{1})-(\ref{8}) are consistent with the heavy-quark
spin-flavor symmetry. The expressions of $\mathcal{A}_{2}$,
$\mathcal{A}_3$ and $\mathcal{B}_2$ are given in the appendix.
Finally, the differential rates of the semileptonic decays of
$B_{s1}$ and $B_{s2}^*$ are

\begin{eqnarray}
&&\frac{d\Gamma}{d\omega}(B_{sJ}\to
D_{sJ'}\ell\bar\nu)=\frac{G_{F}^2|V_{cb}|^2}{(2J+1)48\pi^3}m_{B_{sJ}}r^2(\omega^2-1)^{1/2}\Theta[B_{sJ},
D_{sJ'}]\nonumber\\
\end{eqnarray}
with
\begin{eqnarray}
\Theta[B_{sJ},
D_{sJ'}]=(q_{\mu}q_{\nu}-q^2g_{\mu\nu})W_{J}^{\mu\nu}(D_{sJ'}),
\end{eqnarray}
where

\begin{eqnarray}
&&W_{J}^{\mu\nu}(D_{sJ'})=\sum_{spins}\langle
D_{sJ'}|\gamma^\mu(1-\gamma_5)|B_{sJ}\rangle\langle
D_{sJ'}|\gamma^\nu(1-\gamma_5)|B_{sJ}\rangle^*,
\end{eqnarray}
and $r={m_{D_{sJ'}}}/{m_{B_{sJ}}}$. The allowed integral range for
$\omega$ is
\begin{eqnarray}
0\leq\omega-1\leq
\frac{(m_{B_{sJ}}-m_{D_{sJ'}})^2}{2m_{B_{sJ}}m_{D_{sJ'}}}.
\end{eqnarray}

For the processes of $B_{s1}\to D_{s}\ell\bar{\nu}$, $B_{s1}\to
D_{s}^*\ell\bar{\nu}$, $B_{s1}\to D_{s0}\ell\bar{\nu}$ and
$B_{s1}\to D_{s1}'\ell\bar{\nu}$, the expressions of
$\Theta[B_{s1}, D_{sJ'}]$ are respectively
\begin{eqnarray}
&&\Theta[B_{s1},
D_{s0}(D_s)]\nonumber\\\quad&&=m_{B_{s1}}^4r|\zeta(\omega)|^2\Big\{
\big[ (r\omega-1)f+(\omega^2-1)(rh_1+h_2) \big]^2
+2(r^2-2r\omega+1)[f^2+(\omega^2-1)h_{3}^2]\Big\},\label{1to0}
\end{eqnarray}
\begin{eqnarray}
&&\Theta[B_{s1},
D_{s1}'(D_s^*)]\nonumber\\\quad&&=m_{B_{s1}}^4r|\zeta(\omega)|^2\Big\{
(\omega^2-1)\{[(r\omega-1)f_1-(r-\omega)f_2
+\omega(rh_1+h_2)+(\omega^2-1)h_2']^2+2(rh_1+h_2)^2\nonumber\\&&\quad+
2(1+r^2-2r\omega)(f_1^2+f_2^2)\}
+2[(r\omega-1)h_3+(r-\omega)h_4]^2 +2(1+r^2-2r\omega)\{(\omega
h_3+h_4)^2\nonumber\\&&\quad +[h_3+\omega
h_4+(\omega^2-1)h]^2\}\Big\}.\label{1to1}
\end{eqnarray}
For the semileptonic decays of $B_{s2}^*$, the $\Theta[B_{s1},
D_{sJ'}]$ read
\begin{eqnarray}
&&\Theta[B_{s2}^*,
D_{s0}(D_s)]\nonumber\\\quad&&=m_{B_{s2}}^4r|\zeta(\omega)|^2(\omega^2-1)
\Big\{\frac{2}{3}[(r\omega-1)f+r(\omega^2-1)g]^2
+(1+r^2-2r\omega)[f^2+(\omega^2-1)h^2]\Big\},
\end{eqnarray}
\begin{eqnarray}
&&\Theta[B_{s2}^*,
D_{s1}'(D_{s}^*)]\nonumber\\\quad&&=m_{B_{s2}}^4r|\zeta(\omega)|^2(\omega^2-1)
\Big\{(\omega^2-1)\{\frac{2}{3}[-(r-\omega)f
+\omega(rf_1+f_2)+(r\omega-1)g]^2+(rf_1+f_2)^2\nonumber\\&&\quad+(rh_1+h_2)^2
+(1+r^2-2r\omega)(\frac{4}{3}f^2+g^2+h_1^2 -2h_2^2)\}
+\frac{10}{3}(1+r^2-2r\omega)(h_1+\omega h_2)^2\Big\}.
\end{eqnarray}

\begin{center}
\begin{table}[htb]
\begin{ruledtabular}
\begin{tabular}{c|ccccc}
$B_{s0}\to$&$f$&$h_1$&$h_2$&$h_3$\\\hline

$D_{s}$&-&$1$&$-1$&-&\\
$D_{s0}$&-&$-1$&$-1$&-&\\

$D_{s}^*$&$1-\omega$&$0$&$1$&$1$\\
$D_{s1}'$&$\omega+1$&$0$&$-1$&$1$\\
\end{tabular}
\caption{The coefficients relevant to the transitions of
$B_{s0}\to D_s,D_{s}^*,D_{s0},D_{s1}'$. \label{bs0}}
\end{ruledtabular}
\end{table}
\end{center}

\begin{center}
\begin{table}[htb]
\begin{ruledtabular}
\begin{tabular}{c|ccccccccccc}
$B_{s1}^\prime\to$&$f$&$f_1$&$f_2$&$h_1$&$h_2$&$h_2'$&$h_3$&
$h_4$&$h$\\\hline

$D_{s}$&$\omega-1$&-&-&$-1$&$0$&-&$1$&-\\
$D_{s0}$&$\omega+1$&-&-&$1$&$0$&-&$1$&-&-\\

$D_{s}^*$&-&$-1$&$1$&$1$&$-1$&$0$&$-1$&$1$&$0$\\
$D_{s1}'$&-&$-1$&$-1$&$1$&$1$&$0$&$-1$&$-1$&$0$\\
\end{tabular}
\caption{The coefficients relevant to the transitions of
$B_{s1}^\prime\to D_s,D_{s}^*,D_{s0},D_{s1}'$.\label{bs1p}}
\end{ruledtabular}
\end{table}
\end{center}

In this work, we also calculate the semileptonic decays of the
$B_{s0}$ and $B_{s1}'$ mesons in the $S$ doublet, i.e.
$B_{s0}(B_{s1}')\to D_{s}(D_{s}^*,D_{s0},D_{s1}')\ell\bar\nu$. For
$B_{s0}\to D_{s0}(D_{s1}')\ell\bar\nu$, we have
\begin{eqnarray}
\Theta[B_{s0},D_{s0}]=m_{B_{s0}}^4r|\chi(\omega)|^2(\omega^2-1)(rh_1+h_2)^2,\label{aa1}
\end{eqnarray}
\begin{eqnarray}
&&\Theta[B_{s0},D_{s1}^{\prime}]\nonumber\\&&=m_{B_{s0}}^4r|\chi(\omega)|^2
\{[(\omega-r)f+(\omega^2-1)(rh_1+h_2)]^2
+2(r^2-2r\omega+1)[f^2+(\omega^2-1)h_3^2]\},\label{aa2}
\end{eqnarray}
where
\begin{eqnarray}
\chi(\omega)=Z_S[\mathcal{B}_1+\mathcal{B}_2+m_s]_{\alpha=\Delta_S,\beta=\Delta_S}.
\end{eqnarray}
The $\Theta[B_{s1}^{\prime},D_{s0}]$ and
$\Theta[B_{s1}^{\prime},D_{s1}^{\prime}]$ functions for the
$B_{s1}^{\prime} \to D_{s0}(D_{s1}^{\prime})\ell\bar{\nu}$ decays
are similar to Eqs. (\ref{1to0})-(\ref{1to1}) where
$\zeta(\omega)$ has to be replaced by the new form factor
$\chi(\omega)$. The corresponding parameters and coefficients are
listed in Tables \ref{bs0}-\ref{bs1p}.

For $B_{s0}\to D_{s}(D_{s}^*)l\bar{\nu}$, $\Theta[B_{s0},D_{s}]$
and $\Theta[B_{s0},D_{s}^*]$ can be obtained after replacing
$\chi(\omega)$ by $\lambda(\omega)$ in Eq.
(\ref{aa1})-(\ref{aa2}), where
\begin{eqnarray}
\lambda(\omega)=\sqrt{Z_H
Z_S}[\mathcal{B}_1-\mathcal{B}_2+m_s]_{\alpha=\Delta_S,\beta=\Delta_H}.
\end{eqnarray}
Similarly, one gets the functions $\Theta[B_{s1}',D_{s}]$ and
$\Theta[B_{s1}',D_{s}^*]$ for $B_{s1}'\to
D_{s}(D_{s}^*)\ell\bar{\nu}$ with the replacement
$\zeta(\omega)\to \lambda(\omega)$ in Eqs.
(\ref{1to0})-(\ref{1to1}).

\section{Numerical results}\label{sec4}

We now collect the input parameters: $G_F=1.1664\times 10^{-5}$
GeV$^{-2}$, $V_{cb}=0.043$; $M_{B_{s1}}=5829.4$ MeV,
$M_{B_{s2}^*}=5839.6$ MeV, $M_{D_s}=1968.2$ MeV,
$M_{D^*_s}=2112.0$ MeV, $M_{D_{sJ}^{*}(2317)}=2317.3$ MeV,
$M_{D_{sJ}(2460)}=2458.9$ MeV \cite{PDG}. $M_{B_{s0}}=5718$ MeV
and $M_{B_{s1}'}=5765$ \cite{mass-Bs01}. $m_s=0.5$ GeV,
$\Lambda=1.25$ GeV, the infrared cutoff $\mu=0.593$ GeV and
$\Delta_{S}-\Delta_{H}=335\pm 35$ MeV \cite{parameter}.

In Table \ref{tab}, we give the ranges of $\Delta_{H,S,T}$ for the
strange sector, which are given in Ref.
\cite{parameter,parameter-1}. In terms of the definitions of
$Z_{H,S,T}$ in Eqs. (\ref{ZH})-(\ref{ZT}), we obtain the values of
$Z_{H,S,T}$ listed in Table \ref{tab}.

\begin{center}
\begin{table}[htb]
\begin{ruledtabular}
\begin{tabular}{cccccccc}

Type&    $\Delta_H$ & $\Delta_S$&  $\Delta_T$& $Z_H$  &$Z_S$
&$Z_T$
    \\\hline
(a)&    0.5 & 0.86 &0.84& 4.87 & 2.95& 3.26  \\

(b)&  0.6 & 0.91 &0.94 & 3.45 & 2.28& 1.91  \\

(c)& 0.7 & 0.97 &1.04& 2.37 &1.66& 1.06  \\

\end{tabular}\caption{The values of $\Delta_{H,S,T}$ and the corresponding $Z_{H,S,T}$. Here
$\Delta_{H,S,T}$ and $Z_{H,S,T}$ are in units of GeV and
GeV$^{-1}$, respectively. \label{tab}}
\end{ruledtabular}
\end{table}
\end{center}

\begin{figure}[htb]
\begin{center}
\begin{tabular}{cccc}
\scalebox{0.8}{\includegraphics{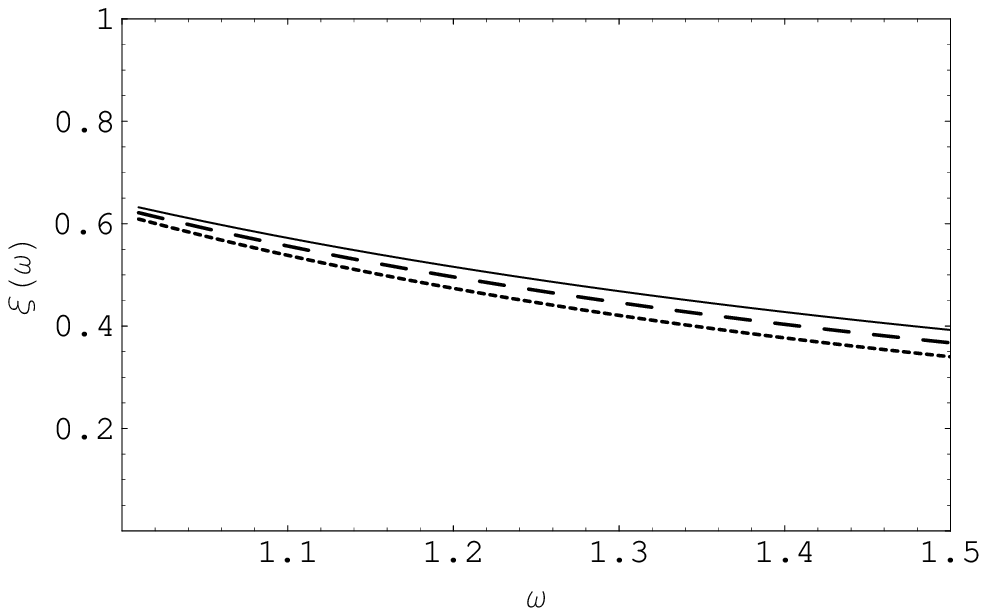}}&\scalebox{0.8}{
\includegraphics{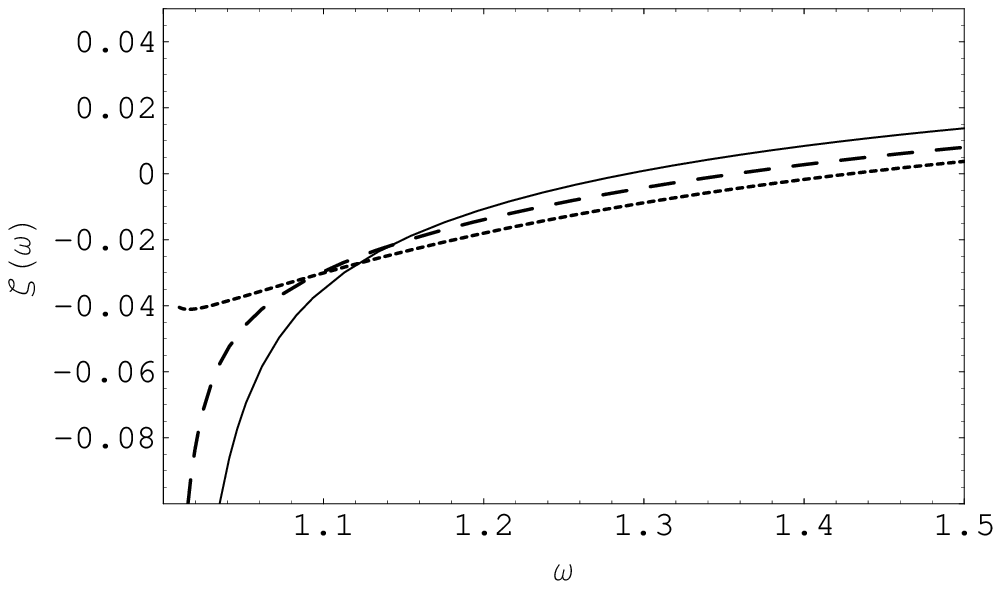}}\\
(a)&(b)\\
\scalebox{0.8}{\includegraphics{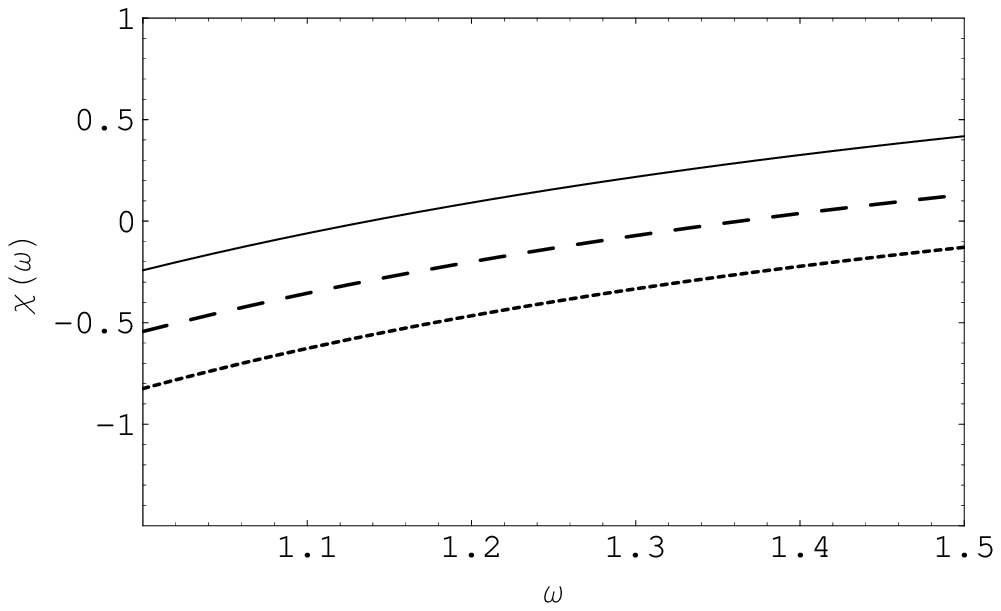}}&\scalebox{0.8}{
\includegraphics{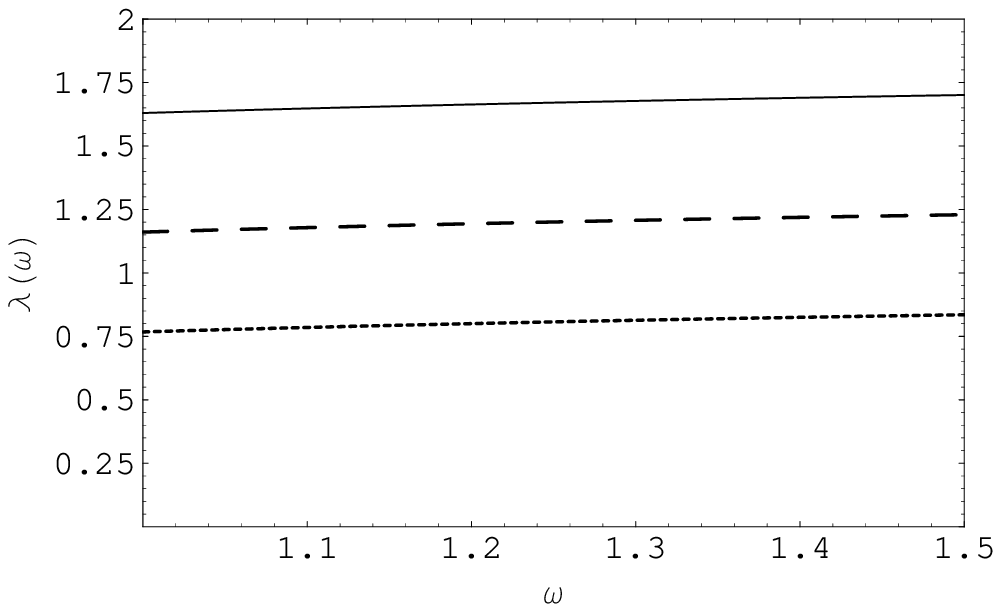}}\\
(c)&(d)
\end{tabular}
\end{center}
\caption{(a) The dependence of $\xi(\omega)$ on $\omega$. Here the
solid, dashed and dotted lines correspond to the results with
parameters $(\Delta_{T}=0.84,\Delta_{H}=0.5)$,
$(\Delta_{T}=0.94,\Delta_{H}=0.6)$,
$(\Delta_{T}=1.04,\Delta_{H}=0.7)$, respectively. (b) The
variation of $\zeta(\omega)$ with $\omega$. Here the solid, dashed
and dotted lines correspond to the results with parameters
$(\Delta_{T}=0.84,\Delta_{S}=0.86)$,
$(\Delta_{T}=0.94,\Delta_{S}=0.91)$,
$(\Delta_{T}=1.04,\Delta_{S}=0.97)$, respectively. (c) The
dependence of $\chi(\omega)$ on $\omega$. Here the solid, dashed
and dotted lines correspond to the results with parameters
$\Delta_{S}=0.86$, $\Delta_{S}=0.91$, $\Delta_{S}=0.97$,
respectively. (d) The variation of $\lambda(\omega)$ with
$\omega$. Here the solid, dashing and dotted lines correspond to
the results with parameters $(\Delta_{S}=0.86,\Delta_{H}=0.5)$,
$(\Delta_{S}=0.91,\Delta_{H}=0.6)$,
$(\Delta_{S}=0.97,\Delta_{H}=0.7)$, respectively.\label{FF}}
\end{figure}

With the above parameters, Fig. \ref{FF} illustrates the
dependence of the form factors $\xi(\omega)$, $\zeta(\omega)$,
$\chi(\omega)$ and $\lambda(\omega)$ on $\omega$.


In Tables \ref{result} and \ref{result2}, we give the decay widths
of the semileptonic decays of $B_{s1}$, $B_{s2}^*$, $B_{s0}$ and
$B_{s1}'$ with the three different choices of parameters
$\Delta_{H,S,T}$ listed in Table \ref{tab}. Up to now, the CDF and
D0 collaborations have not measured the decay widths of $B_{s1}$
and $B_{s2}^*$.

\begin{center}
\begin{table}[h]
\begin{ruledtabular}
\begin{tabular}{c||ccc|ccc}
&&$B_{s1}\to$&&&$B_{s2}^*\to$&\\\cline{2-7}
&(a)&(b)&(c)&(a)&(b)&(c)\\\cline{1-7}

$D_{s}\ell\bar{\nu}$&$2.1\times10^{-15}$&$1.8\times10^{-15}$&$1.6\times10^{-15}$&
$2.1\times10^{-15}$&$1.8\times10^{-15}$&$1.6\times10^{-15}$\\

$D_{s}^*\ell\bar{\nu}$&$4.9\times10^{-15}$&$4.4\times10^{-15}$&$3.9\times10^{-15}$&
$5.0\times10^{-15}$&$4.5\times10^{-15}$&$3.9\times10^{-15}$\\

$D_{s0}\ell\bar{\nu}$&$8.7\times10^{-20}$&$4.7\times10^{-20}$&$6.9\times10^{-20}$&
$5.6\times10^{-20}$&$3.0\times10^{-20}$&$4.5\times10^{-20}$\\

$D_{s1}'\ell\bar{\nu}$&$1.0\times10^{-19}$&$8.7\times10^{-20}$
&$1.5\times10^{-19}$&$1.2\times10^{-19}$&$1.0\times10^{-19}$&$1.8\times10^{-19}$\\
\end{tabular}\caption{The decay widths of the
semileptonic decays of $B_{s1}$ and $B_{s2}^*$. Here columns (a),
(b), (c) correspond to the results with the three parameter
combinations listed in Table \ref{tab}. The decay widths are in
units of GeV. \label{result}}
\end{ruledtabular}
\end{table}
\end{center}
\begin{center}
\begin{table}[h]
\begin{ruledtabular}
\begin{tabular}{c||ccc|ccc}
&&$B_{s0}\to$&&&$B_{s1}^{'}\to$&\\\cline{2-7}
&(a)&(b)&(c)&(a)&(b)&(c)\\\cline{1-7}

$D_{s}\ell\bar{\nu}$&$2.5\times10^{-14}$&$1.3\times10^{-14}$&$5.9\times10^{-15}$&$1.2\times10^{-14}$&$6.4
\times10^{-15}$&$2.9\times10^{-15}$\\

$D_{s}^*\ell\bar{\nu}$&$2.5\times
10^{-15}$&$1.3\times10^{-14}$&$5.9\times10^{-15}$&$3.8\times10^{-14}$&$2.0\times10^{-14}$&$9.1\times10^{-15}$\\

$D_{s0}\ell\bar{\nu}$&$1.6\times10^{-15}$&$3.9\times10^{-16}$&$2.8\times10^{-15}$&$4.9\times10^{-16}$&$9.6
\times10^{-16}$&$3.7\times10^{-15}$\\

$D_{s1}'\ell\bar{\nu}$&$1.5\times10^{-15}$&$3.3\times10^{-15}$&$1.3\times10^{-14}$&
$2.1\times10^{-15}$&$2.7\times10^{-15}$&$1.2\times10^{-14}$\\

\end{tabular}\caption{The decay widths of the
semileptonic decays of $B_{s0}$ and $B_{s1}^{\prime}$. Here
columns (a), (b), (c) correspond to the results with the three
parameter combinations listed in Table \ref{tab}. The decay widths
are in units of GeV. \label{result2}}
\end{ruledtabular}
\end{table}
\end{center}

\begin{widetext}
\begin{center}
\begin{figure}[htb]
\begin{tabular}{cc}
\scalebox{0.7}{\includegraphics{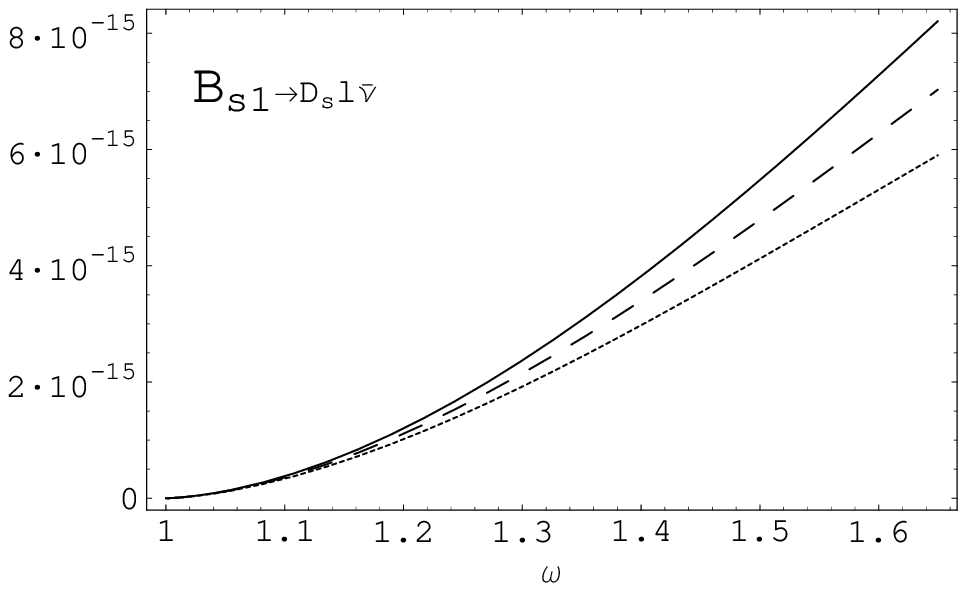}}&
\scalebox{0.7}{\includegraphics{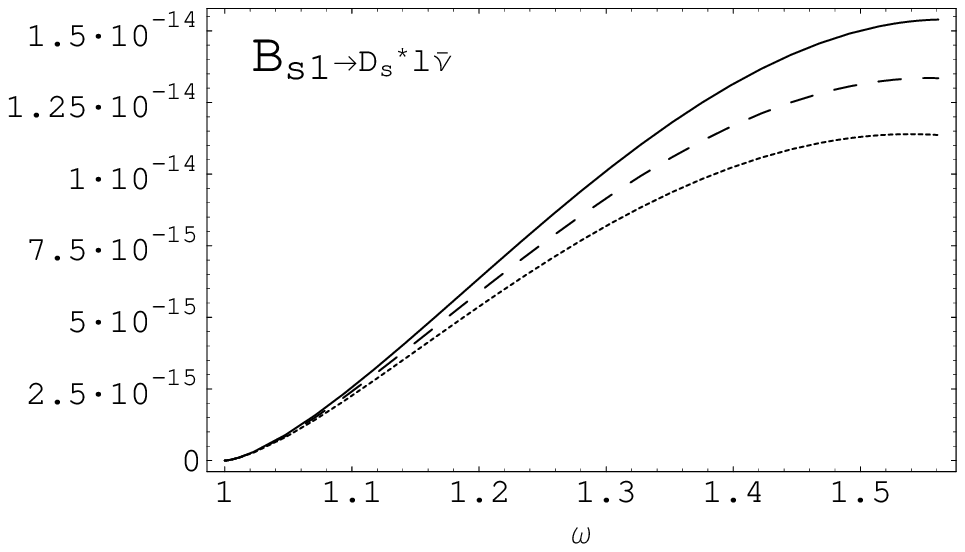}}\\
(a)&(b)\\
\scalebox{0.7}{\includegraphics{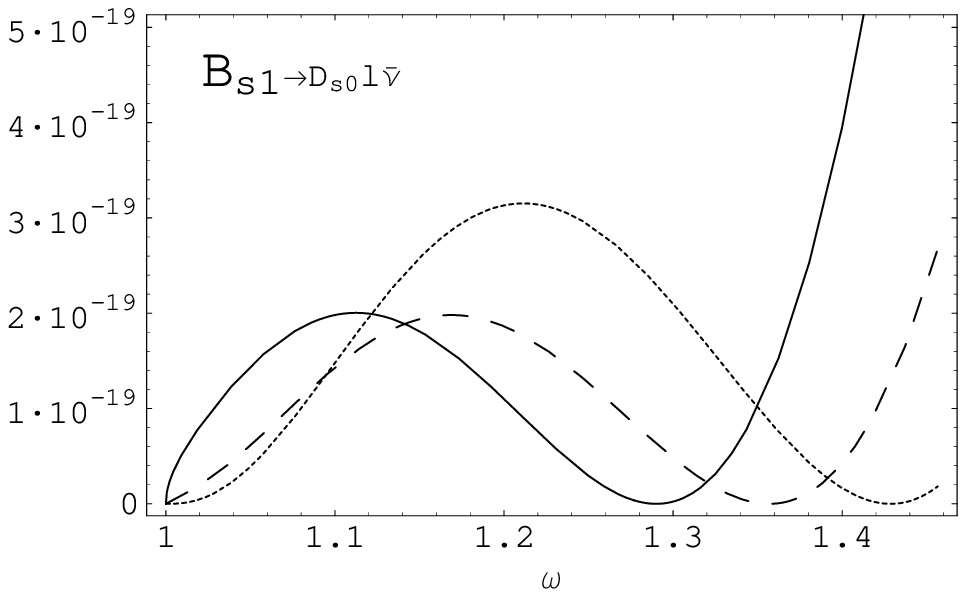}}&\scalebox{0.7}{\includegraphics{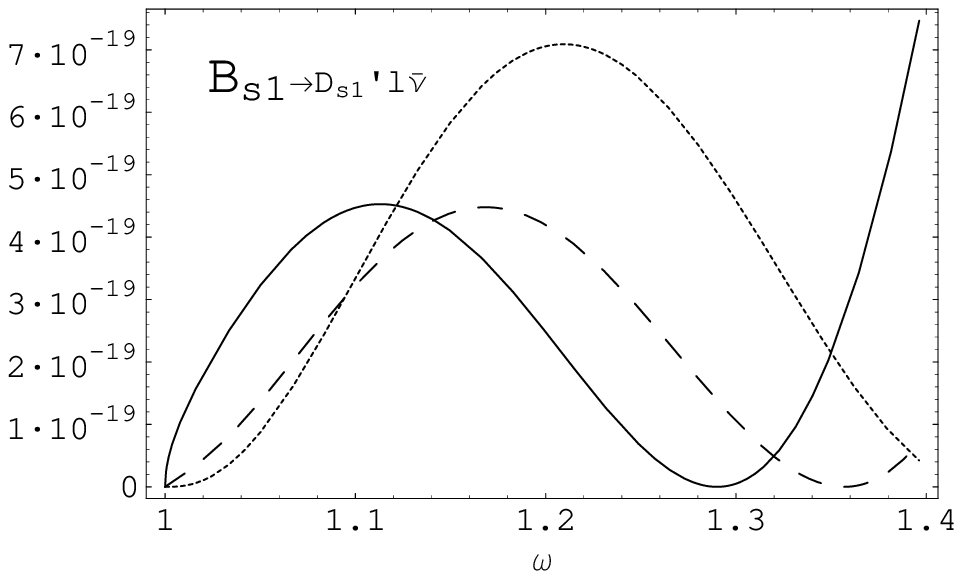}}\\(c)&(d)\\
\end{tabular}
\caption{(a), (b), (c) and (d) illustrate the dependence of the
differential rates of the semileptonic decays of $B_{s1}$ on
$\omega$. Here the solid, dashed  and dotted lines correspond to
type (a), (b) and (c) in Table \ref{tab}.\label{partial}}
\end{figure}
\end{center}
\end{widetext}

Using the semileptonic decays of $B_{s1}$ as an example, we show
the variation of the differential rates of the semileptonic decays
of $B_{s1}$ with $\omega$ in Fig. \ref{partial}. The differential
rates of $B_{s1}\to D_{s}\ell\bar{\nu}$ and $B_{s1}\to
D_{s}^*\ell\bar{\nu}$ increase with $\omega$ monotonically. In
contrast, the variation of the differential rates of $B_{s1}\to
D_{s0}\ell\bar{\nu}$ and $B_{s1}\to D_{s1}^\prime \ell\bar{\nu}$
with $\omega$ are not monotonic. The different line shapes of the
differential rates of the semileptonic decays of $B_{s1}$ account
for the results of the decay widths in Table \ref{result}, i.e. it
tells us why the decay rates for type (c) are not always larger
than the results for the other types. The same observation holds
for the semileptonic decays of $B_{s2}^*$, $B_{s0}$,
$B_{s1}^\prime$.

\section{Discussion}\label{sec5}

The semileptonic decay of the bottom-strange meson is an
interesting topic. Due to the lack of experimental information of
the excited bottom-strange mesons, theorists mainly focused on the
semileptonic decay of $B_{s}$ using theoretical approaches such as
the QCD sum rule approach \cite{qsr} and the quark model
\cite{xiangliu-2}. The observations of $B_{s1}$ and $B_{s2}^*$
\cite{CDF,D0} enrich the mass spectrum of the bottom-strange
system greatly. In this work, in order to explore the possibility
of searching the semileptonic decays of the excited bottom-strange
states in the experiments, we have calculated
$B_{s0},\,B_{s1}^{\prime},\,B_{s1},\,B_{s2}^*\to
[D_s(1968),\,D_{s}^*(2112),\,D_{sJ}(2317),\,
D_{sJ}(2460)]\ell\bar{\nu}$ in the framework of the CQM model.

Our numerical results indicate that (1) the decay width of
$B_{s1}(B_{s2}^*)\to D_{s}(D_{s}^*)\ell\bar{\nu}$ is around
$10^{-15}$ GeV, which is $4\sim5$ orders of magnitude larger than
that of $B_{s1}(B_{s2}^*)\to D_{sJ}(2317,2460)\ell\bar{\nu}$; (2)
the decay width of $B_{s0}(B_{s1}')\to D_{s}(D_{s}^*,
D_{sJ}(2317,2460))\ell\bar{\nu}$ is $10^{-14}\sim 10^{-15}$ GeV.

Although the CDF and D0 experiments did not measure the decay
widths of $B_{s1}$ and $B_{s2}^*$ and the experiments did not
observe $B_{s0}$ and $B_{s1}^{\prime}$, we can roughly estimate
the branching ratio of the semileptonic decay of $B_{s1}$,
$B_{s2}$, $B_{s0}$ and $B_{s1}^{\prime}$. In Ref. \cite{liu-3P0},
one obtains the two-body strong decay widths of $B_{s1}$ and
$B_{s2}^*$ as 98 keV and 5 MeV respectively. $B_{s0}$ and
$B_{s1}'$ are expected to be narrow resonances with a width around
several tens of keV\footnote{In Ref. \cite{lujie}, authors
calculated the isospin-violating strong decay widths of $B_{s0}$
and $B_{s1}^\prime$, which are 35 keV and 38 keV, respectively. }
\cite{lujie}, since their main decay modes are the isospin
violating strong decays and electromagnetic decays. Thus it is
reasonable to take the strong decay width as the total width
approximately for $B_{s1}$, $B_{s2}$, $B_{s0}$ and
$B_{s1}^{\prime}$. We further show the order of magnitude of the
semileptonic decay of $B_{s1}$, $B_{s2}$, $B_{s0}$ and
$B_{s1}^{\prime}$ in Table \ref{br}, which is convenient for the
experimentalist to conclude whether the current and future
experiments can reach these semileptonic decays.
\begin{center}
\begin{table}[htb]
\begin{ruledtabular}
\begin{tabular}{c||cccccc}
&$B_{s0}\to$&$B_{s1}^{\prime}\to$&$B_{s1}\to$&$B_{s2}^*\to$\\\hline

$D_{s}\ell\bar{\nu}$&$10^{-9}\sim 10^{-10}$&$10^{-9}\sim
10^{-10}$&$\sim10^{-11}$&
$\sim10^{-13}$\\

$D_{s}^*\ell\bar{\nu}$&$10^{-9}\sim 10^{-10}$&$10^{-9}\sim
10^{-10}$&$\sim10^{-11}$&
$\sim10^{-13}$\\

$D_{s0}\ell\bar{\nu}$&$10^{-10}\sim 10^{-11}$&$10^{-10}\sim
10^{-11}$&$\sim10^{-16}$&
$\sim10^{-18}$\\

$D_{s1}'\ell\bar{\nu}$&$10^{-9}\sim 10^{-10}$&$10^{-9}\sim
10^{-10}$&$\sim10^{-16}$&
$\sim10^{-17}$\\
\end{tabular}\caption{The estimation of the branching fractions of the
semileptonic decays of $B_{s0}$, $B_{s1}^{\prime}$, $B_{s1}$ and
$B_{s2}^*$ according to our numerical result shown in Table
\ref{result} and \ref{result2}. \label{br}}
\end{ruledtabular}
\end{table}
\end{center}

From Table \ref{br}, we can exclude the possibility of finding the
semileptonic decay of $B_{s2}^*\to
[D_{s},\,D_{s}^*,\,D_{s0},\,D_{s1}']\ell\bar{\nu}$ and $B_{s1}\to
[D_{s0},\,D_{s1}']\ell\bar{\nu}$ in experiments. However, for
$B_{s1}\to [D_{s},\,D_{s}^*]\ell\bar{\nu}$ and the semileptonic
decays of $B_{s0}$ and $B_{s1}^\prime$, the upper limit of the
branching ratio can reach to $10^{-9}$. The present precision of
the experimental measurement of the branching fraction of the $B$
mesons has reached up to $10^{-7}\sim 10^{-8}$ \cite{PDG}. The
decays $B_{s1}\to [D_{s},\,D_{s}^*]\ell\bar{\nu}$ and the
semileptonic decays of $B_{s0}$ and $B_{s1}^\prime$ may be
observed in future experiments. Especially, the forthcoming LHCb
experiments will produce an enormous amount of data of
heavy-flavor hadrons, which is one of the potential experiments in
which to search the $B_{s1}\to [D_{s},\,D_{s}^*]\ell\bar{\nu}$ and
the semileptonic decays of $B_{s0}$ and $B_{s1}^\prime$. Then
these semileptonic decays will be helpful to further test the
structure of $D_{sJ}(2317)$ and $D_{sJ}(2460)$. In our
calculation, we have assumed $D_{sJ}(2317)$ and $D_{sJ}(2460)$ as
the charm-strange mesons with $J^P=0^+$ and $1^+$. If the
experimental measurement of these semileptonic decays are
consistent with our prediction, it will provide strong support of
the $c\bar{s}$ structure for $D_{sJ}(2317)$ and $D_{sJ}(2460)$.

\section*{Acknowledgments}

This project was supported by the National Natural Science
Foundation of China under Grants 10625521, 10721063, 10705001.
X.L. was also supported by the \emph{Funda\c{c}\~{a}o para a
Ci\^{e}ncia e a Tecnologia of the Minist\'{e}rio da Ci\^{e}ncia,
Tecnologia e Ensino Superior} of Portugal (SFRH/BPD/34819/2007).

\section*{Appendix}

The definitions of $\mathcal{I}_{0}(\alpha)$, $\mathcal{I}_1$,
$\mathcal{I}_5(\alpha,\beta,\omega)$,
$\mathcal{I}_6(\alpha,\beta,\omega)$, $\mathcal{I}_{3}(\alpha)$,
$\mathcal{A}_2(\alpha,\beta,\omega)$,
$\mathcal{A}_3(\alpha,\beta,\omega)$,$\mathcal{B}_1(\alpha,\beta.\omega)$
and $\mathcal{B}_2(\alpha,\beta,\omega)$ are \cite{CQM}

\begin{eqnarray}
&&\mathcal{A}_2(\alpha,\beta,\omega)\nonumber\\
\quad&&=[(\omega^2-1)T(\alpha,\beta,\omega)-6\omega U(\alpha,\beta,\omega)\nonumber\\
&&\quad+(2\omega^2+1)S(\alpha,\beta,\omega)+3S(\beta,\alpha,\omega)]/[2(\omega^2-1)^2],\nonumber\\\\
&&\mathcal{A}_3(\alpha,\beta,\omega)\nonumber\\
\quad&&=[-\omega(\omega^2-1)T(\alpha,\beta,\omega)+2(2\omega^2+1)U(\alpha,\beta,\omega)\nonumber\\
&&\quad-3\omega(S(\alpha,\beta,\omega)+S(\beta,\alpha,\omega))]/[2(\omega^2-1)^2]
\end{eqnarray}

\begin{eqnarray}
\mathcal{B}_1(\alpha,\beta,\omega)=[\omega
X(\beta,\alpha,\omega)-X(\alpha,\beta,\omega)]/[\omega^2-1],
\end{eqnarray}

\begin{eqnarray}
\mathcal{B}_2(\alpha,\beta,\omega)=[\omega
X(\alpha,\beta,\omega)-X(\beta,\alpha,\omega)]/[\omega^2-1],
\end{eqnarray}
with
\begin{eqnarray}
T(\alpha,\beta,\omega)=\mathcal{I}_6(\alpha,\beta,\omega)+m_s^2\mathcal{I}_5(\alpha,\beta,\omega),
\end{eqnarray}
\begin{eqnarray}
U(\alpha,\beta,\omega)=\mathcal{I}_1+\alpha \mathcal{I}_3(\alpha)
+\beta\mathcal{I}_3(\beta)+\alpha\beta\mathcal{I}_5(\alpha,\beta,\omega),
\end{eqnarray}
\begin{eqnarray}
S(\alpha,\beta,\omega)=\omega[\mathcal{I}_1+\beta\mathcal{I}_3(\beta)]
+\alpha\mathcal{I}_3(\beta)+\alpha^2\mathcal{I}_5(\alpha,\beta,\omega),\nonumber\\
\end{eqnarray}
\begin{eqnarray}
X(\alpha,\beta,\omega)=-\mathcal{I}_3(\beta)-\alpha\mathcal{I}_5(\alpha,\beta,\omega),
\end{eqnarray}
\begin{eqnarray}
\mathcal{I}_{0}(\alpha)&=&\frac{N_c}{16\pi^{3/2}}\int_{1/\Lambda^2}^{1/\mu^2}\frac{dy}{y^{3/2}}e^{-y(m_s^2-\alpha^2)}
\left(\frac{3}{2y}+m_s^2-\alpha^2\right)\nonumber\\&&\times\left[1+\mathrm{erf}(\alpha\sqrt{y})\right]-\alpha\frac{N_c
m_{s}^2}{16\pi^2}
\Gamma\left(-1,\frac{m_s^2}{\Lambda^2},\frac{m_s^2}{\mu^2}\right),\nonumber\\
\end{eqnarray}
\begin{eqnarray}
\mathcal{I}_{1}&=&\frac{N_c
m_s^2}{16\pi^2}\Gamma\left(-1,\frac{m_s^2}{\Lambda^2},\frac{m_s^2}{\mu^2}\right),
\end{eqnarray}
\begin{eqnarray}
\mathcal{I}_{3}(\alpha)&=&\frac{N_c}{16\pi^{3/2}}\int^{1/\mu^2}_{1/\Lambda^2}\frac{dy}{y^{3/2}}
\exp[-y(m_{s}^{2}-\alpha^2)]\left(1+\mathrm{erf}(\alpha\sqrt{y})\right),
\end{eqnarray}
\begin{eqnarray}
&&\mathcal{I}_5(\alpha,\beta,\omega)\nonumber\\\quad&&=\int_0^1dx\frac{1}{1+2x^2(1-\omega)+2x(\omega-1)}
\Big[\frac{2N_c}{16\pi^{3/2}}\int_{1/\Lambda^2}^{1/\mu^2}dy\sigma
e^{-y(m_s^2-\sigma^2)}y^{-1/2}(1+\mathrm{erf}(\sigma\sqrt{y}))\nonumber\\&&+\frac{2N_c}{16\pi^2}
\int_{1/\Lambda^2}^{1/\mu^2}dy e^{-y m_s^2}y^{-1}\Big],\nonumber\\\\
&&\mathcal{I}_6(\alpha,\beta,\omega)\nonumber\\\quad&&=\mathcal{I}_1
\int_{0}^{1}dx\frac{\sigma}{1+2x^2(1-\omega)+2x(\omega-1)}
-\frac{N_c}{16\pi^{3/2}}\int_{0}^{1}dx\frac{1}{1+2x^2(1-\omega)+2x(\omega-1)}\nonumber\\&&\quad
\times\int_{1/\Lambda^2}^{1/\mu^2}dy
y^{-3/2}e^{-y(m_s^2-\sigma^2)}\Big\{\sigma[1+\mathrm{erf}(\sigma\sqrt{y})]
[1+2y(m_s^2-\sigma^2)]+2\sqrt{\frac{y}{\pi}}\left[\frac{3}{2y}+(m_s^2-\sigma^2)\right]\Big\},\nonumber\\
\end{eqnarray}
where
$$\sigma(\alpha,\beta,\omega)=\frac{(1-x)\alpha+x\beta}{\sqrt{1+2x^2(1-\omega)+2x(\omega-1)}}.$$


\end{document}